\documentclass[12pt]{iopart}
\usepackage{graphicx}

\begin{document}

\title{Relaxation of creep strain in paper}
\author{Mika Mustalahti, Jari Rosti, Juha Koivisto, Mikko J. Alava}
\address{Aalto University, Department of Applied Physics,
FIN-00076 Aalto, Finland}
\ead{mika.mustalahti@helsinki.fi}
\ead{jari.rosti@tkk.fi}
\ead{mjalava@cc.hut.fi}
\ead{juha.koivisto@tkk.fi}





\begin{abstract}
In disordered, viscoelastic or viscoplastic materials a sample
response exhibits a recovery phenomenon after the removal of a
constant load or after creep. We study experimentally the recovery in
paper, a quasi two-dimensional system with intrinsic structural
disorder. The deformation is measured
by using the digital image correlation (DIC) method. By the DIC we
obtain accurate displacement data and the spatial fields of deformation and recovered strains.
The averaged results are first compared to
several heuristic models for in particular viscoelastic polymer materials.
The most important experimental quantity is the permanent creep strain,
and we analyze whether it is non-zero by fitting the empirical models
of viscoelasticity.
We then present in more detail the spatial
recovery behavior results from DIC, and show that they indicate a power-law -type relaxation.
We outline results on sample-to-sample
variation and collective, spatial fluctuations in the recovery behaviour.
An interpretation is provided of the relaxation in the general
context of glassy, interacting systems with barriers.
\end{abstract}

\maketitle

\date{\today}

\section{Introduction}

The nonlinear behavior of materials is highly topical thanks to
several observations of "universality", that is generic behavior
that is not dependent on the particulars of the example at hand.
Examples for such generality can be found from the crackling noise
or intermittent response by which many systems - magnets (Barkhausen
noise), dislocation assemblies (strain avalanches), foams, granular
systems, fracture and creep deformation etc. - react to external
forces or influences
\cite{stefano,zaiser,weiss,UCH-04,Advphys,MIG-01}. Likewise,
universal behavior is found in "aging" and "rejuvenation", two
concepts first introduced in the theory of spin glasses where at the
simplest these can be taken to imply that the correlation or
response properties of the system are found to be dependent on the
observation or waiting time scale \cite{jpb}. This dependency is
moreover often non-linear.

The interest in these questions has many origins. On one hand, one
can look for classes of behavior - as in crackling noise or in the
aging scenarios, where in both cases simple models lead the way by
giving examples of distinct behaviors. On other hand, for a
materials scientist or an engineer it is important to gain access to
reliable effective descriptions of materials or "laws" and equations
that summarize the nonlinearities. In particular, for viscoelastic
and viscoplastic polymers this has meant the accumulation of vast
quantities of semi-empirical viscoelastic theories and associated
measurement results \cite{vebook}. For creep, or material deformation
under a constant load, there are also similar empirical bodies of
data and "master equations" to explain their generic features a
classical example being logarithmic creep
deformation. There have been recent advances in treating 
"interacting elastic systems" in such creep conditions
in statistical mechanics, in the presence of disorder 
(\cite{creep} and see also \cite{carmen}
for a connection to viscoplastic deformation).

To combine the fundamental physical issues and practical (but often
much more extensive) observations is one challenge for the
statistical mechanics of materials deformation. In this work, we
study the recovery of creep strain in paper in this context. Fiber networks
(as paper) consist of a "frozen" random structure which changes only
as a result of microscopic fracture damage and energy dissipation
related to irreversible plastic deformation
\cite{AlavaNiskanen}. On the smaller scale, fibers are
viscoelastic-plastic "beams" that themselves exhibit creep behavior
if tested individually. The fact that the typical creep behavior of
the network does not equal that of single fibers highlights the
importance of coarse-grained (or "collective") behavior
\cite{AlavaNiskanen,santucci}. Studies of the physics of paper have
noted that paper has the interesting rheological property
of delayed recovery. Following a stress-strain cycle or creep load,
the remaining stress at zero strain is time-dependent
(\cite{AlavaNiskanen, Niskanen}). Similar mechanical phenomena 
exist in other "fibrous networks" that have received recently attention
such as bucky- and graphene oxide paper and actine networks
\cite{bucky,graphene,actine}. In polymeric systems it has been suggested 
that the rheology is controlled by the motion of defects \cite{polymer}. 
Extremely long term creep-recovery studies have been also conducted for 
polymers \cite{polymer2}. 
In the case of crystalline materials, polycrystalline ice has been demonstrated
upon unloading to show logarithmic delayed deformation
\cite{duval}. This is fairly close to what we find below.

\begin{figure}[ht!]
\centering
\includegraphics[width=10cm]{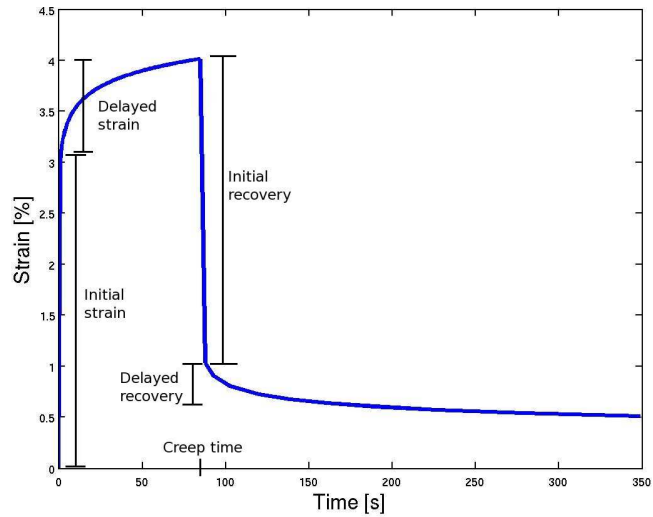}
\caption{A schematic strain-time curve of a creep-recovery experiment
with the initial strain, the delayed strain, the initial recovery
and the delayed recovery marked in. The sum of the initial strain and the
delayed strain is called the total strain and the sum of the initial recovery
and the delayed recovery is called the total recovered strain.}
\label{fig:creeprecoverysample}
\end{figure}

Here, we study the dynamics of strain recovery after the
loading through one creep cycle. A
sample is in each test creep-strained at a constant stress and then
the strain decrease and the deformation field is measured as a
function of time. A typical creep-recovery curve is presented in
Figure \ref{fig:creeprecoverysample}. In the creep-recovery the
sample experiences first an initial recovery and after that delayed
recovery follows. In the delayed recovery the rate of recovery
decreases until the sample has recovered all the possible deformation
(in paper physics, for a discussion of these issues see 
\cite{Brezinski, Coffin}).
The essential questions are thus two: i) what is the non-recoverable
strain if any, and ii) what is the dependence of the relaxation on
time? A further question is what can be stated of microscopic dynamics 
and its relation to the sample-level, coarse-grained response.

By the usage of the Digital Image Correlation -technique (DIC) we
can extract with very high accuracy the global sample
strain and also the spatial strain fields, thus also the details of the
creep and recovery phases \cite{hild, bausch}.
The main results that we gain are the following (Figure 
\ref{fig:creeprecoverysample} presents the definitions of the relevant 
quantities). First of all, by
careful measurements by almost five orders of magnitude in time, we gain
information on the ensemble-averaged behavior of the recovery. This
can be contrasted with empirical models of viscoelasticity, 
and we also find that part of the creep strain remains permanent. 
Certain models work fairly in comparison with the data, while most are
clearly not valid. It appears that the relaxation can be best fitted
by a power-law with a small exponent, with - as stated
already - a remnanent strain in the limit of infinite times.
We study the sample-to-sample and spatial variations of the initial
creep part and the corresponding recovery dynamics. The
spatial recovery rates follow a distribution, whose width
scales with time in the same manner as the average recovery rate.
This indicates an asymmetry compared to the dynamics under
loading, which means that there is no simple ``superposition principle''
relating the two dynamics under loading and recovery.

The structure of this paper is as follows. In Section 2, we present
the experimental details as far as the procedures and the DIC
technique are concerned. In Section 3, we show the experimental
data, and discuss some heuristic models and their comparison with
the data. We also present a possible interpretation of the data via
an energy landscape picture. This allows also to consider the role
of internal stresses in the relaxation. Section 4 finishes the paper with a
discussion.

\section{Methods}

\subsection{Experimental setup}

\begin{figure}[ht!]
\centering
\includegraphics[width=.5\textwidth]{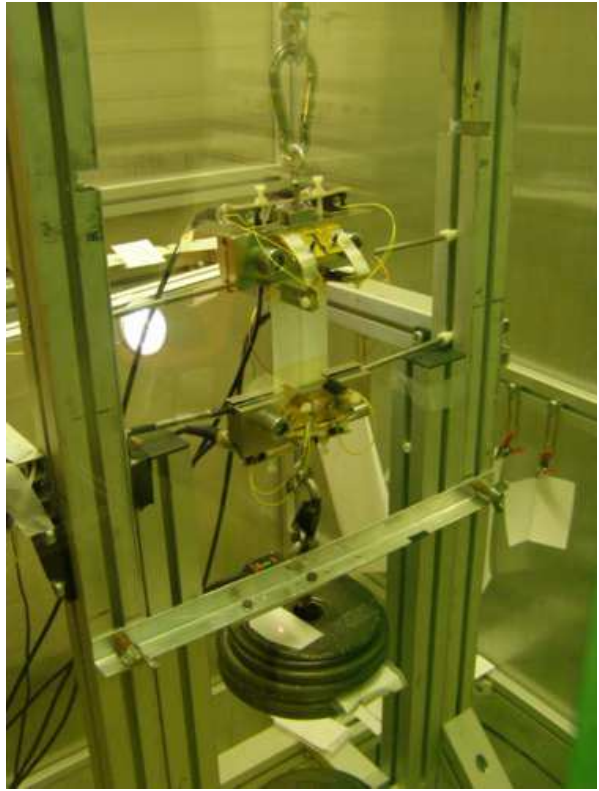}
\caption{The setup for creep experiments.  A load is attached to the lower clamp and its movement is controlled by using pneumatic cylinders. During the unloading phase the pneumatic cylinders let the sample recover freely without any load. When pictures are taken during the recovery, the load due to the weight of the lower clamp, is temporarily applied to the sample. }
\label{fig:setup}
\end{figure}

Pictures are taken of the sample during the creep and the recovery process and the displacement field is measured by using the digital image correlation method.
The setup is shown in Figure \ref{fig:setup}. A 100~mm long and
30~mm wide sample is attached to the  upper and lower clamps. 
The upper clamp is attached
directly to the frame while the lower clamp with a mass of 855~g is
let hang freely. The load is attached to the lower clamp, which can
be moved to up and down positions using a pneumatic cylinder. 
The speed of the lift movement is set to 1~cm/s, 
so that the load is applied steadily. 
In its unloaded state the paper has an 1~cm slack, and the load is fully 
applied within
order of one second after the lift movement has started.

A global displacement measurement device, Omron laser distance sensor, is attached to the measurement device to follow the movement of the lower clamp.  The results of the global displacement measurement is used for the consistency checking of DIC results, since its accuracy was found to be 0.1~mm due to mechanical limitations and the accuracy of the sensor.

\begin{figure}[ht!]
\centering
\includegraphics[width=.5\textwidth]{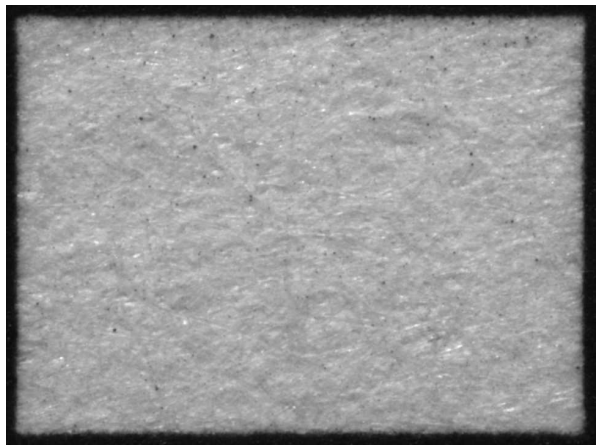}
\caption{A fibre scale image from paper. The image size is approximately 4x3 mm and there are about 950x700 pixels. The image is taken from the centre of the sample.}
\label{fig:image}
\end{figure}

The sample was imaged with PCO's 1 mega pixel grayscale digital
camera, SensiCam 370 KL 0562. This camera has a very low thermal noise
ratio. An example of the images taken is in Figure \ref{fig:image}.
Images were taken at an exponentially decreasing rate during the recovery:
18 images during the first hour and 18 pictures after one hour  of recovery.

The lower clamp has a mass of 855~g, and since the paper thickness is 0.1mm,
the stress applied to the sample during imaging was 2.8~MPa.
The sample was loaded fo 30 to 60 seconds during the recovery phase due to the imaging.
The total elastic strain due to loading was $0.13\pm0.01\%$ since an elastic modulus $2.2\pm0.1$~GPa was measured. Strain due to lower clamp mass was well below yield strain $0.3\%$ and the order $0.1\%$ which is regarded as a linear viscoelastic limit of paper in engineering: in Ref.~\cite{Lif} it is shown that if the stress applied to paper is below 0.1\%, nearly all of the deformation caused by the stress is recovered immediately after the stress is removed. The relative strain caused by the load application during imaging is of the order of error bars presented in the Fig.~\ref{fig:creep-recovery14} and the disturbances due to the load application to the displacement measurement are smaller. 

\subsection{Digital Image Correlation}

The digital image correlation is the task of finding a deformation function, mapping coordinates
from a reference image to coordinates in the test image \cite{kybic,hild,sutton}. The deformation function is presented as cubic splines, where knots are defined in an evenly spaced grid, with knot spacing h$\times$h pixels (crate).  The exact algorithm for the deformation computation is described in
\cite{kybic2}. A cubic spline approximation of the deformation function
leads to a locally minimized elastic energy and 
it can represent global affine deformations
correctly. 
The method lies between ``global'' and ``local'' methods. In the global
approach one defines a single criterion, e.g. affine deformation, which is
globally optimized. In the case of local methods, one minimizes the
error in each zone of interest, and a global criterion
for the deformation function is not defined.

In our analysis a crate size of 64 pixels was used. The
boundary errors were minimized by printing black frames to the sample, and aiming the camera to the centre of the frames. The black
colour printed with a regular office printer was uniform enough to
prevent the algorithm from seeing any texture in it.

\begin{figure}
\begin{minipage}[t]{.45\textwidth} 
\centering
\includegraphics[width=.9\textwidth]{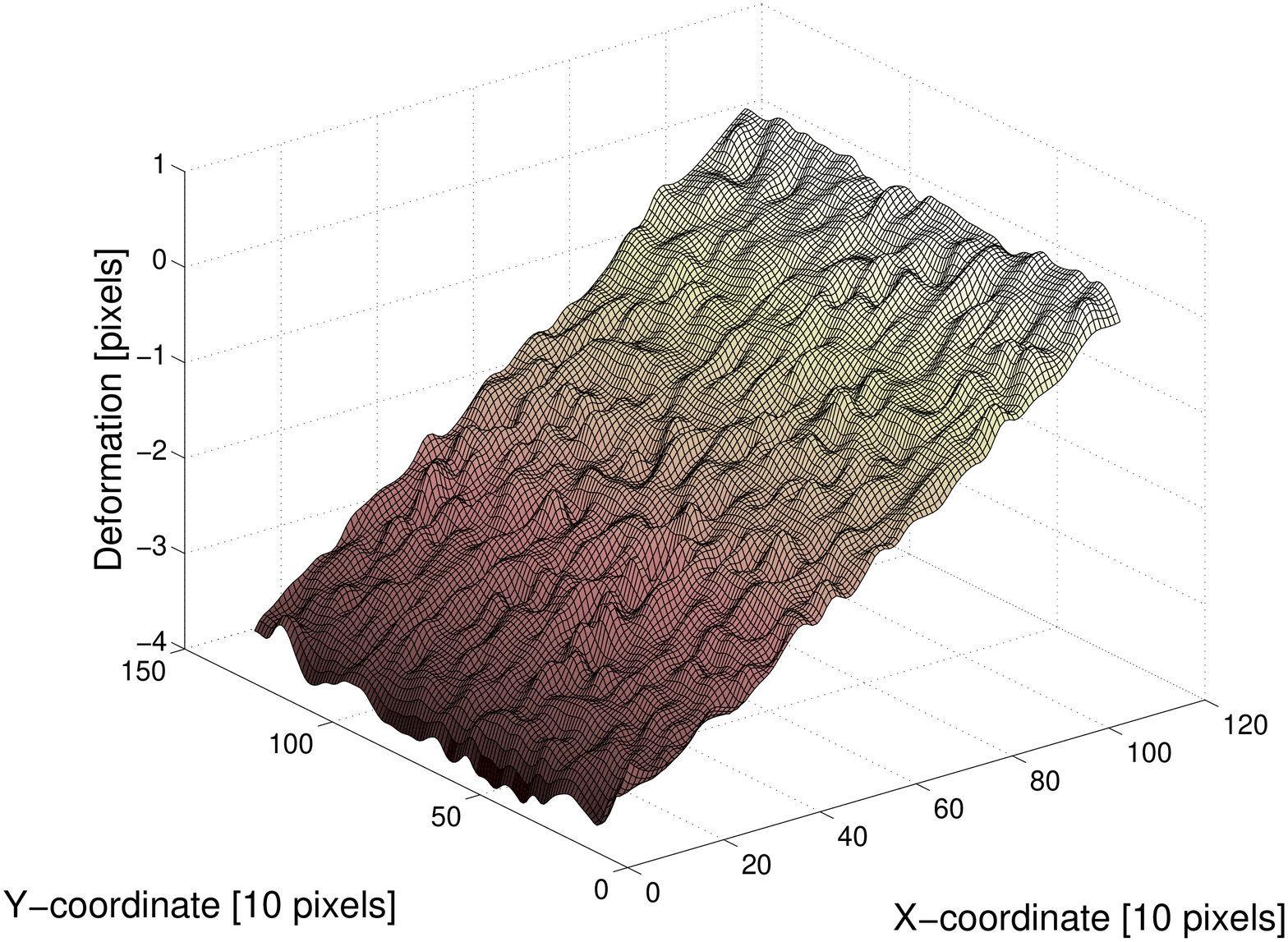}
\caption{Image deformed with artificially made linear deformation and a small bump. The crate size of the computation is 16 pixels.}
\label{fig:accuracyc16deflin}
\end{minipage}
\hspace{.05\textwidth} 
\begin{minipage}[t]{.45\textwidth}
\centering
\includegraphics[width=.9\textwidth]{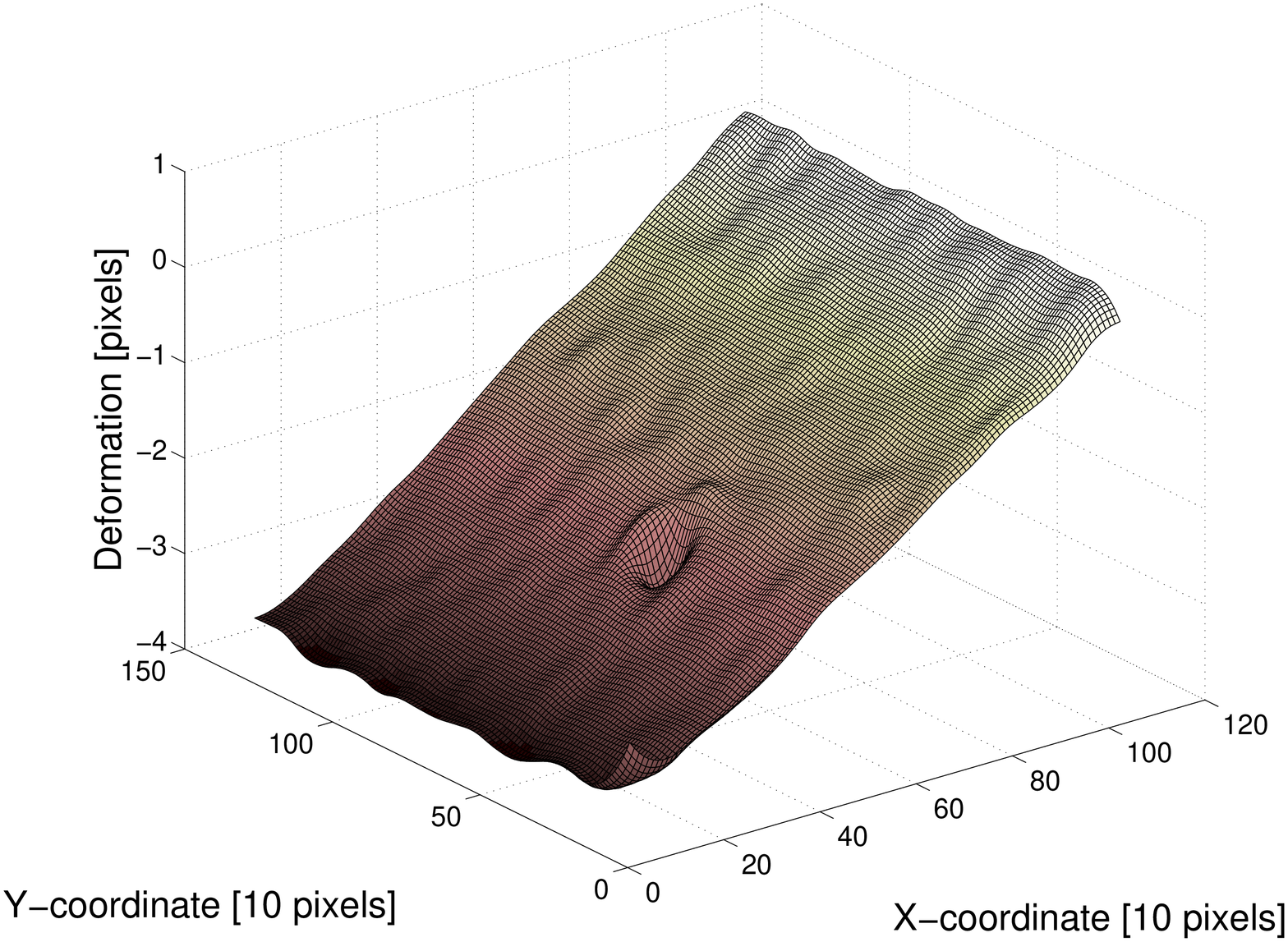}
\caption{Image deformed with artificially made linear deformation and a small bump. The crate size of the computation is 32 pixels. }
\label{fig:accuracyc32deflin}
\end{minipage}
\end{figure}

The crate 64x64 pixels was chosen based on tests of the algorithm.
For example, an artificial deformation resembling the experimental one was made to the image. 
The displacement fields computed with crate sizes of 16 and 32 pixels are in Figures \ref{fig:accuracyc16deflin} and \ref{fig:accuracyc32deflin},
respectively. With both crate sizes, the linear displacement is clearly seen but in the Fig.~\ref{fig:accuracyc32deflin}, crate size of 32 pixels, the displacement field is much smoother than in the Figure \ref{fig:accuracyc16deflin} where a crate size of 16 pixels was used.
The artificial deformation was found and a correct amount of deformation was obtained, but in the area where the deformation should be linear, an average error of about 0.2 pixels is seen.
After a further change of the crate size the noise created by the algorithm is reduced significantly, being less than 0.1~pixels.
Increasing the crate size improves the accuracy, but reduces the details 
seen in the spatial displacement field.
In order to measure the strain accurately, we used a rather large
64x64 pixel crate size, since the main purpose was to measure the
global strain. In Ref. \cite{kybicthesis} there is
additional discussion about the accuracy of the method.

\subsection{Strain measurement}

We measure the strain in experiments using two different methods:
the global displacement on the image area and the spatial
average of spatial strain rates. When the global displacement of the image area is used,  we take an initial picture as a reference image and compare all consecutive images to the initial one. The global displacement between initial and the reference is measured using equation
\begin{equation}
d = \langle y'-y \rangle_x,
\label{eq:dist}
\end{equation}
where global displacement on the image area $d$ is averaged over the $x$ 
positions.


For the spatial average of the local strain rate we define an evenly spaced grid on an image, which consists of a discrete set of points $\{(x_1, y_1), (x_2, y_2), ...\}$. The test and reference images are defined as consecutive images taken during the experiment. A spatial strain rate $\dot{\epsilon}^{i,j}$ in a grid point $(i,j)$ is computed from
\begin{equation}
\dot{\epsilon}^{i,j} = \frac{\Delta y}{\Delta t L},
\label{eq:locstrain}
\end{equation}
where $\Delta y$ is defined as an y-directional displacement of grid point $(i,j)$. $\Delta t$ is the time difference between images. The distance to the lower clamp is computed from fitting the plane to displacements and estimating the position where the displacement is zero. The spatial average of the strain rate is computed over the grid points, to then yield the recovery
strain rate during the experiment.

\subsection{Measurements}

We performed 32 first creep first recovery experiments.  The creep times in
the experiments were mainly between 20 and
500 seconds in length. In two experiments the creep time was longer: 1230 and
3078 seconds.

The creep in paper, delayed deformation depicted in
Fig.~\ref{fig:creeprecoverysample}, can be divided, as for many materials, into three stages in time, primary, secondary and tertiary creep \cite{Coffin}. 
When the load is applied to the paper the primary creep (delayed strain) 
starts immediately. In the primary creep the strain follows Andrade's power 
law at relatively high stresses
\begin{equation}
 \epsilon(t) = a\hspace{1mm}t^b + c,
\end{equation}
where $a$, $b$ and $c$ are constants and $t$ is the creep time \cite{Brezinski}. In paper, the rate of creep does not exhibit secondary regime where the creep rate is constant, but rather the creep rate 
continues to decrease until the tertiary regime \cite{Coffin}. 
Rheological phases are sometimes defined so that the deformation gained in the primary creep is  recoverable, and the secondary creep deformation is  nonrecoverable \cite{Brezinski, Coffin}.  
Tertiary creep begins roughly when the strain rate starts to increase 
after reaching its minimum at the end of the secondary creep. 
After that, the strain rate accelerates until the sample fails.

We chose a wide range of total creep strains as a starting point to 
the recovery.
We observed a primary creep close to Andrade's law varying from sample
to sample as $\epsilon(t) 
\sim t^{0.1 ... 0.3}$, so that the exponent $b$ was smaller than the 
$b=1/3$ associated typically to Andrade creep.
A sharp transition to secondary creep was not observed, but the exponent values $b$ had a
tendency to increase slowly during the experiment.

The sample material used in the experiments was ordinary
copy paper with a basis weight of 80~g/m$^2$. As usual, industrial
paper has an anisotropy which is denoted by the so-called Machine
Direction (MD) and Cross Direction (CD): if strained in the
principal directions they differ e.g. in the typical degree of
ductility in a tensile test, where the CD turns to be much more
ductile.

In the creep phase of every experiment a load with a mass of 4071~g
was applied to the sample in the CD. With this load we get an initial
stress of 13.3~MPa. Environmental conditions were kept at
constant $33\pm1$ \% of relative humidity and $33\pm1$ $^o$C of
temperature.  With the parameters used, the absolute humidity of the
environment is equal to the situation, where the relative humidity is 50
\% and temperature 23~$^o$C, which is the standard
operating point of mechanical testing of paper. The studies of 
Ketoja et al. showed that the master curve analysis - for 
logarithmic creep - can be used by introducing a moisture 
-dependent shift coefficient. They observed for the creep rate an 
exponential dependence, similarly to the elastic modulus of paper \cite{ketoja}. The fact that the dependence of the viscoplastic
deformation on environmental conditions can be "scaled away" makes
us expect that qualitatively the relaxation would not change as a function of the moisture.

\section{Results}

\subsection{Overview}

\begin{figure}[t]
\centering
\includegraphics[width=.7\textwidth]{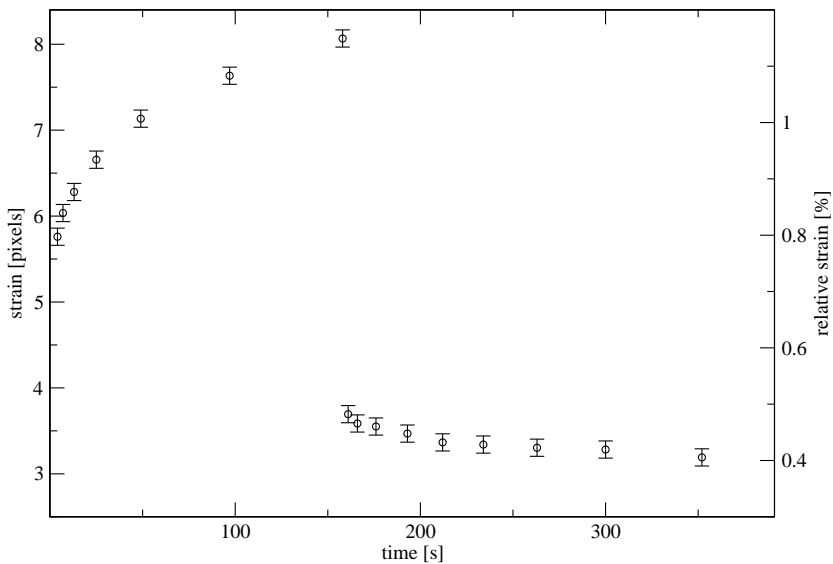}
\caption{A typical example of the deformation curve in a
creep-recovery experiment.  The total strain is computed as the
total displacement of the image area to the strain direction using a
displacement field obtained from DIC using Eq. (\ref{eq:dist}). Error
bars indicate a 0.1 subpixel limit. Pixel size is 4~$\mu$m. In the
creep regime the sample deforms according to Andrade's law  with an
exponent $\epsilon(t) \sim t^{0.1}$. } \label{fig:creep-recovery14}
\end{figure}

In Figure~\ref{fig:creep-recovery14} there is  a typical strain curve
from one creep-recovery experiment. The strain was measured approximately 3
seconds after the load was applied. The load was applied to the sample at
$t=0$ and the sample responded with a strain of approximately 0.55~\% which
consists of instantaenous initial strain and creep which starts to accumulate
immediately when the load is applied to the sample.

The initial strain is followed by a delayed strain that increases
until the load is removed. In the measurement shown in Figure
\ref{fig:creep-recovery14} the load was removed after 150 seconds of
creep. Immediately after the removal, the sample underwent
initial recovery which was  estimated by measuring the strain approximately
3 seconds after the removal of the load. In Figure~\ref{fig:creep-recovery14} the initial recovery
is roughly 0.7\%. After the initial recovery the recovery process of the sample
continues, and the part taking place after the initial recovery is called the
delayed recovery.

\subsection{Averaged creep recovery}

Next we will present the averaged time-dependent recovery, and compare it
to a number of heuristic models. These models start with the typical
linear viscoelasticity-like Maxwell-Kelvin one, which consists of simple elements
describing an ideal viscous and an ideal elastic material. It has the form
\begin{equation}
\epsilon(t)=ae^{-\frac{t}{b}}+d,
\label{eq:maxwell-kelvin-fit}
\end{equation}
where $a$, $b$ and $d$ can be seen as independent fitting
parameters. The parameter $d$ expresses the amount of unrecoverable
strain. Similarly the sum of the parameters $a$ and $d$ is the
amount of the remaining strain immediately after the initial recovery.

A stretched exponential model (also motivated by the
Fancey model of viscoelastic response, \cite{Fancey, Fancey2}) is also used,
\begin{equation}
\epsilon(t)=ae^{-\left( \frac{t}{b} \right)^c}+d,
\label{eq:fanceyfit}
\end{equation}
where $a$, $b$, $c$ and $d$ are independent fitting parameters.
Similarly to the Maxwell-Kelvin model also here the last parameter
$d$ relates to the amount of unrecoverable strain and the sum of $a$
and $d$ should be the  remaining strain immediately after the 
initial recovery.

Another heuristic model is Schapery's thermodynamical model, where
stresses were independent state variables and the entropy production and the
Gibbs free energy were specially defined \cite{Schapery2}:
\begin{equation}
\epsilon(t)=a \left[ \left( 1+b \frac{t-t^{'}}{t^{'}} \right)^c -
\left( b \frac{t-t^{'}}{t^{'}} \right)^c \right],
\label{eq:schaperyfit}
\end{equation}
where $a$, $b$ and $c$ are independent fitting parameters. The
parameter $t^{'}$ is the creep time of the sample which in this case
was 592 seconds. The parameter $a$ tells us again the amount of strain
left immediately after the initial recovery.

Finally, we also tried the combination of a power-law decay and a
constant,
\begin{equation}
 \epsilon(t)=(at)^{-b}+d,
\label{eq:pl}
\end{equation}
where $a$, $b$ and $d$ are independent fitting parameters. The
parameter $d$ is again the unrecoverable strain.

\begin{figure}[t]
\centering
\includegraphics[width=.7\textwidth]{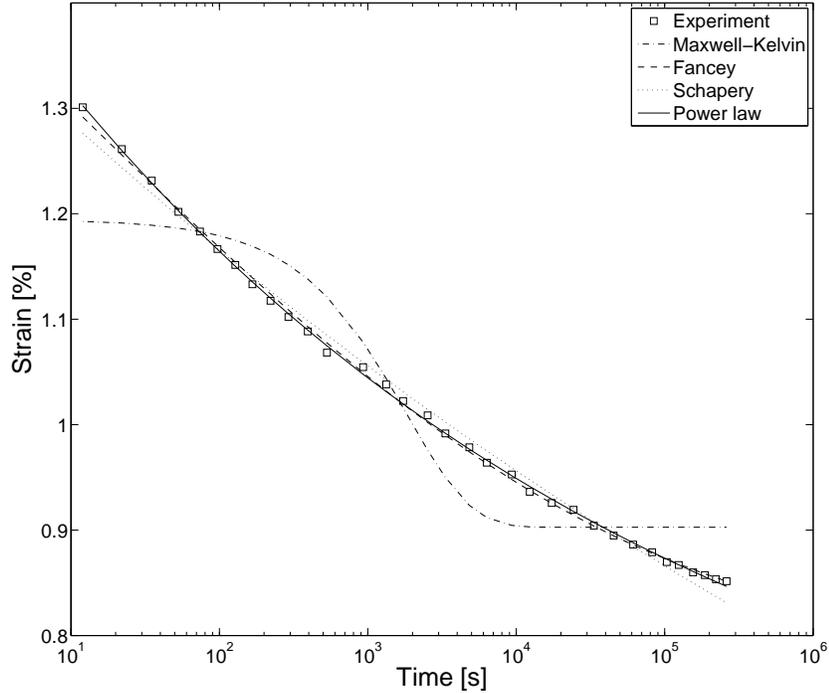}
\caption{Averaged recovery as a function of the recovery time. Four model expressions fitted to the experimental recovery data. The recovery is obtained by comparing the image taken right after the initial recovery to later images using Eq. (\ref{eq:dist}). Experimental data is average over 32 experiments.}
\label{fig:modelscompared}
\end{figure}

In Figure \ref{fig:modelscompared} the four cases are tried with
fits to the experimental recovery data. The data is computed using
global displacement on the image area using Eq. (\ref{eq:dist}).

The first, the Maxwell-Kelvin model is clearly proven insufficient. Fancey's
stretched exponential model is slightly better than Schapery's
thermodynamical model, and finally the power law fit is at least as
good as Fancey's model. However, a decisive conclusion between an
exponential or a power law is difficult.

The fitting parameters for the cases in Figure
\ref{fig:modelscompared} are shown in Table \ref{tab:fitparameters}.
From the Table we conclude that the 
unrecoverable strain does not exhibit as much variation from model to model 
as the strain after initial recovery.
However the unrecoverable strain values from the various fits are still
quite different. Therefore, much longer experiments would have been
necessary to estimate the real asymptotic behavior of the recovery. To half the 
recoverable strain would take roughly 500 times longer, according to the 
Fancey-style fit (Eq.~(\ref{eq:fanceyfit})).
The power-law fit exponent (0.11) could be taken to correspond to a
logarithmic decay, as an analogy to
logarithmic creep. However, the role of fluctuations
becomes in logarithmic creep more important with time, in contrast to what
happens in relaxation which we discuss below. 

Only three parameters in Table \ref{tab:fitparameters} have their
errors estimates presented because only those parameters are
remaining nearly constant for all of the individual recovery curves. 
All of the other parameters are more or less related to the sample properties
which vary from sample to sample. In other words the three parameters
relate to the general shape of the recovery curve.

\begin{table}
\begin{center}
\begin{tabular}[b]{|l| c| c | c| c| c| c|}
\hline
& a [\%] & b & c & d [\%] & US & SAIR \\
\hline
{\small Maxwell-Kelvin} & 0.29 & 1830 s & & 0.90 & 0.90 & 1.19 \\
\hline
Fancey & 1.46 & 5.82 s & 0.10 $\pm$ 0.02 & 0.79 & 0.79 & 2.25 \\
\hline
Schapery & 1.61 & 3730 & 0.96 $\pm$ 0.01 & & & 1.61 \\
\hline
Power law & 2.19 & 0.11 $\pm$ 0.02 & & 0.58 & 0.58 & \\
\hline
\end{tabular}
\end{center}
\caption{Fitting parameters for different models. The amount of unrecoverable strain (US)
 and strain after initial recovery (SAIR) obtained from the fitting parameters are in the rightmost columns.}
\label{tab:fitparameters}
\end{table}

\subsection{Spatial deformation fields}

\begin{figure}
\centering
\includegraphics[width=.32\textwidth]{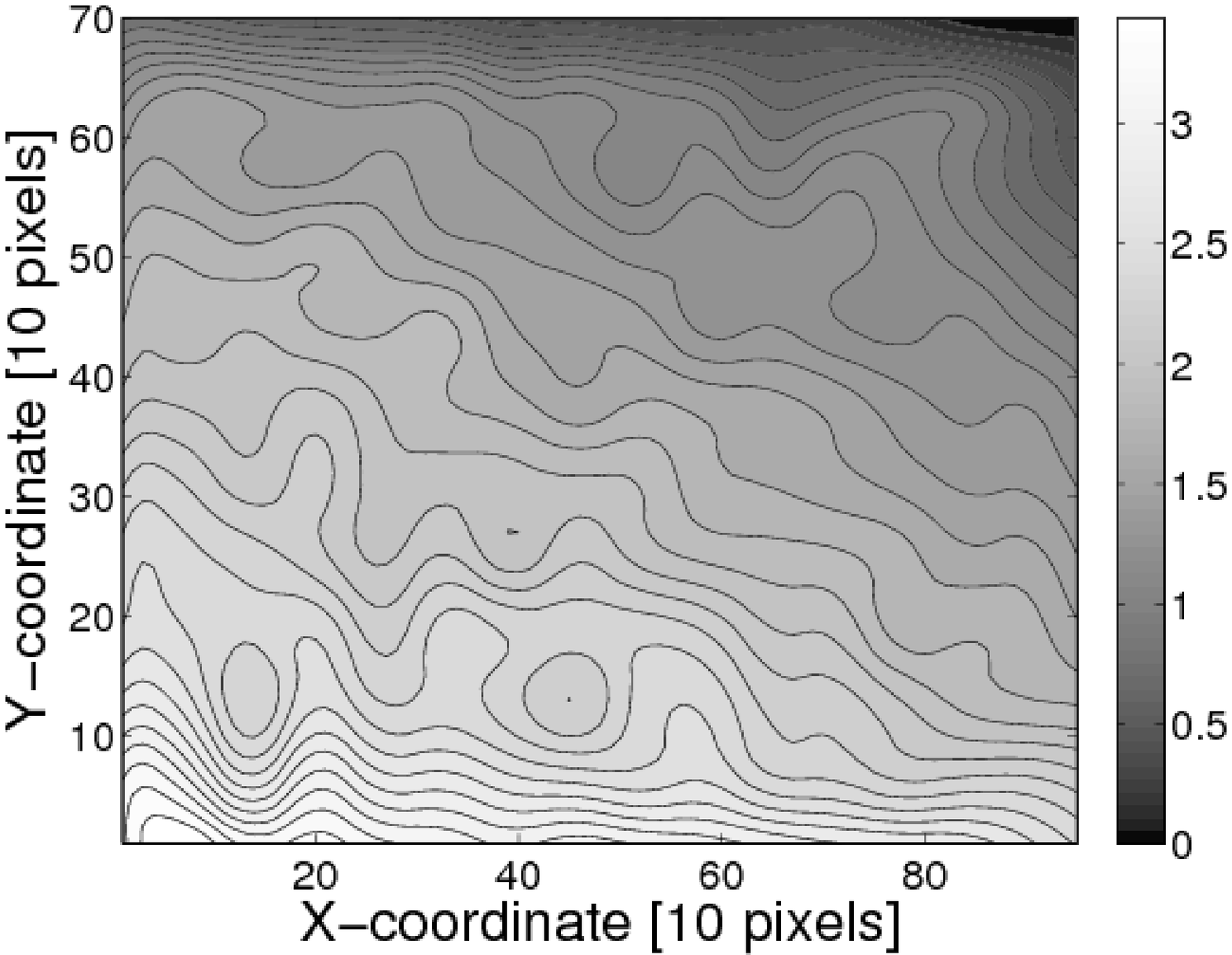}
\hfill
\includegraphics[width=.32\textwidth]{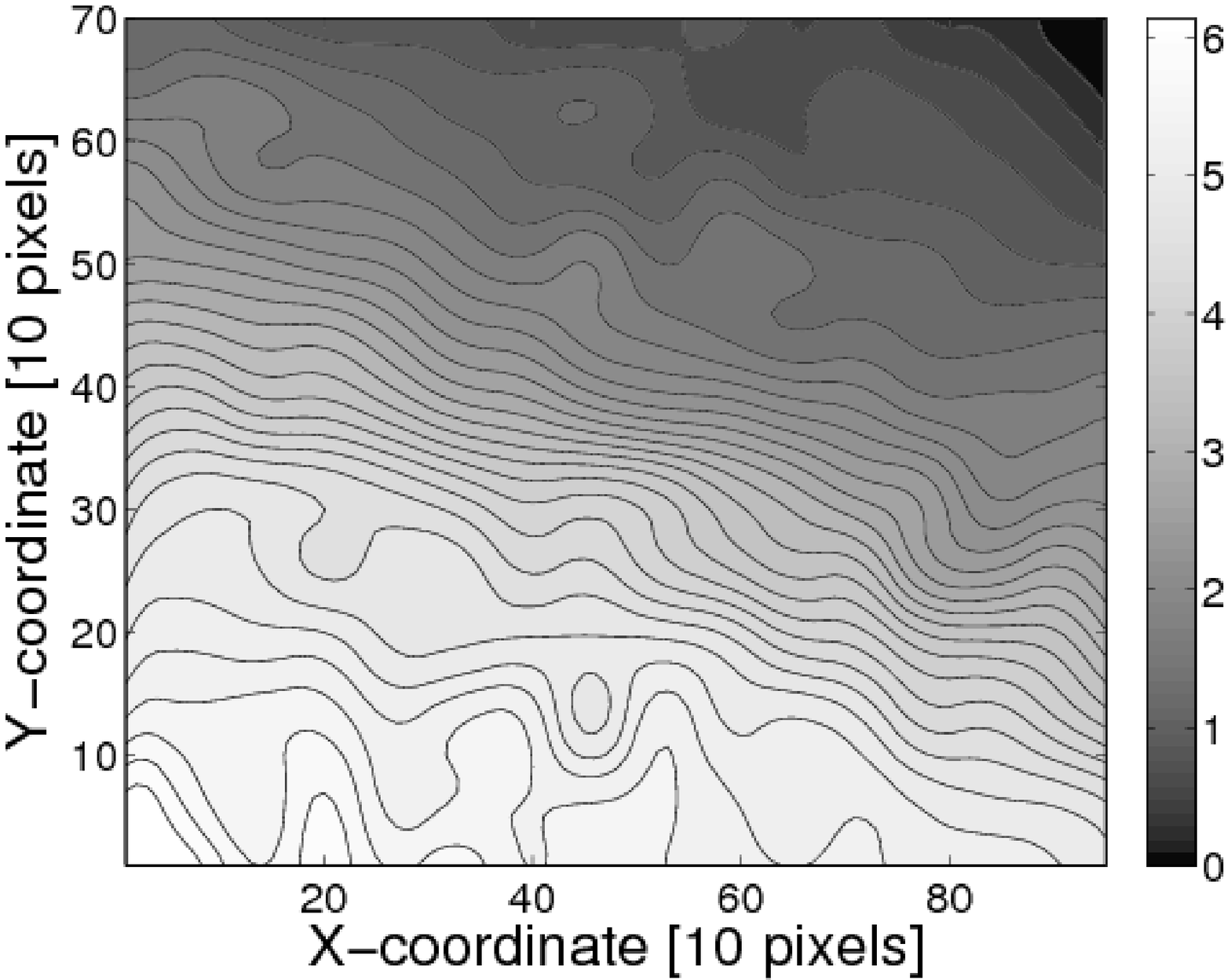}
\hfill
\includegraphics[width=.32\textwidth]{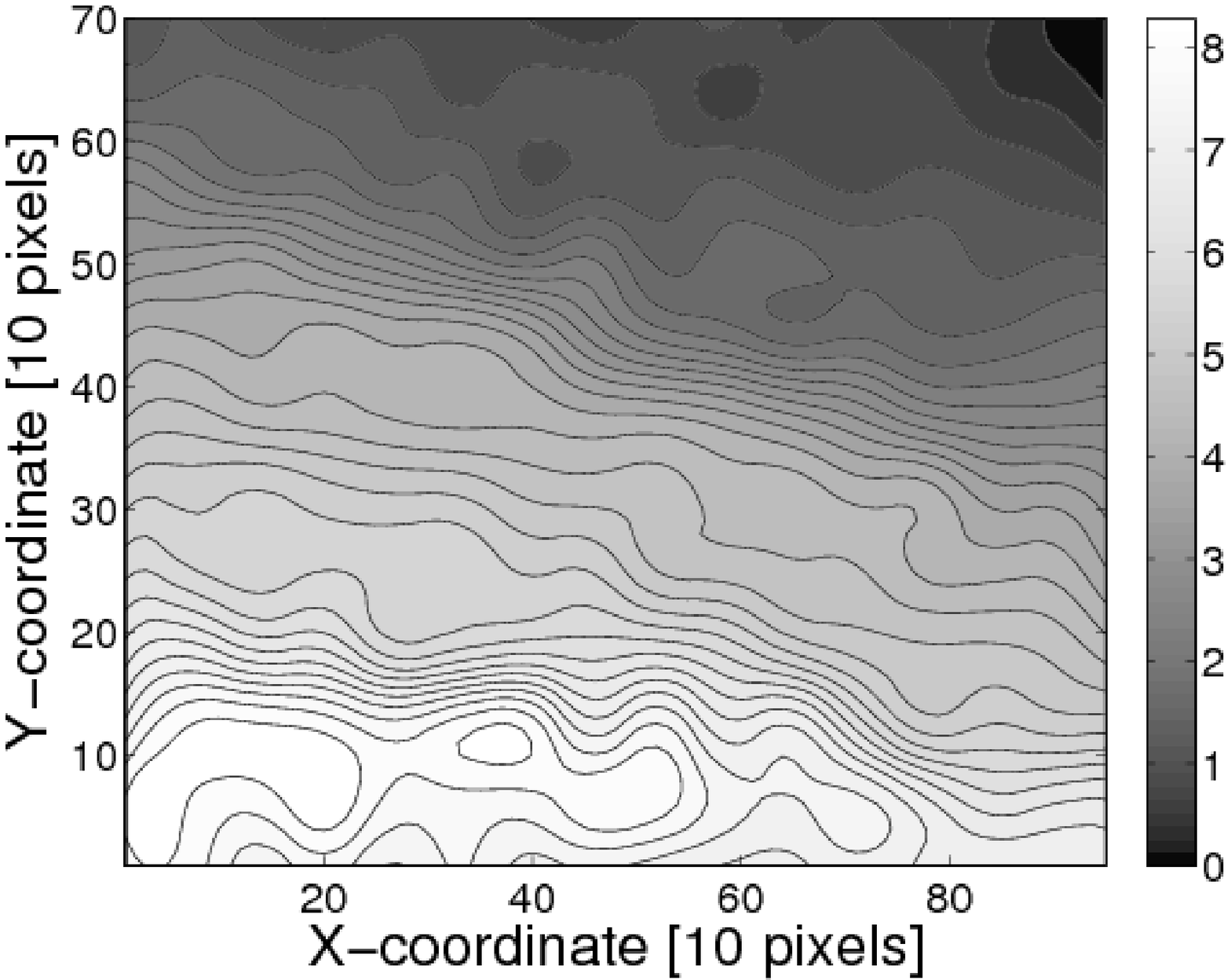}
\caption{The displacement fields after 30, 1700 and 823000 seconds of delayed recovery, respectively. Total amount of a local recovered strain is shown in micrometers in the color-bar. }
\label{fig:g3852kg36delrecovery1}
\end{figure}

The process of delayed recovery is illustrated in Fig.~\ref{fig:g3852kg36delrecovery1}. The
displacement fields in these figures are calculated by comparing the image
taken right after the initial recovery to the later images. The displacement is presented in micrometers.
The total amount of spatial, recovered strain can be read from the colorbar next to each displacement field.
From Fig.~\ref{fig:g3852kg36delrecovery1} it can be seen that the displacement field is
heterogeneous during the delayed recovery. Such local dynamics
in the displacement field  continue upto the end of the experiment. The heterogenity of
the displacement field is obvious, but the experimental setup does not allow us to draw very detailed
conclusions for two reasons. The first is that our image area is limited to
only a small part of the recovering sample, and the second reason is that we are integrating recovery over
timescales which differ over several magnitudes. 
Thus we limit the consideration of the spatial deformation field to the essential properties 
of strain rate distributions and compare those
to an analogous study done for primary creep \cite{puolikas}.

\begin{figure}[t]
\includegraphics[width=.7\textwidth]{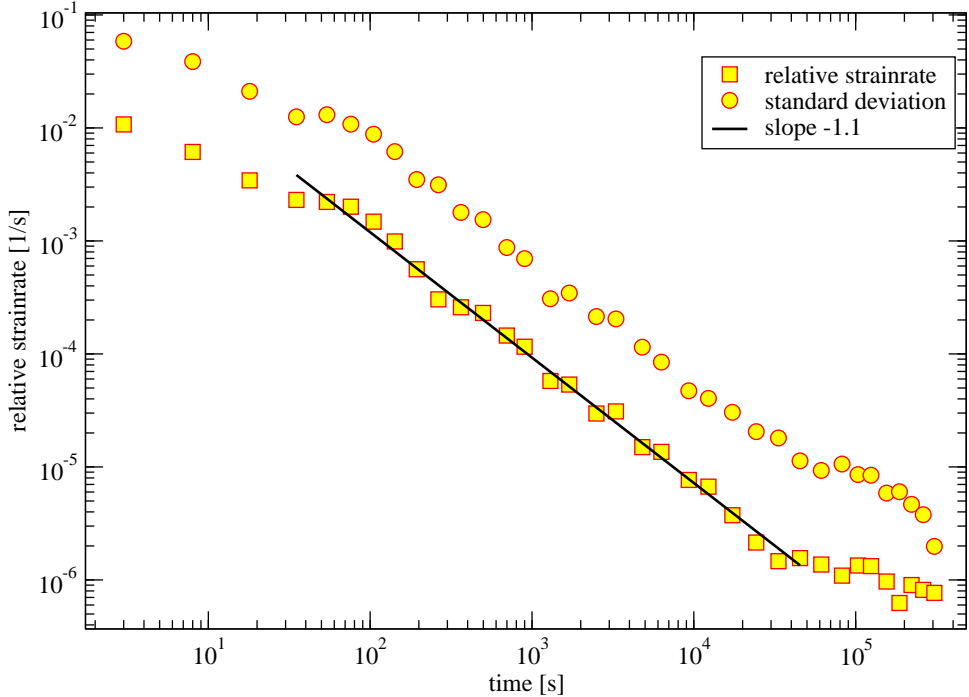}
 \caption{Averaged recovery rate and the standard deviation of spatial recovery rates as a function of recovery time. The data is averaged over 32 experiments. The recovery rate agrees with the global recovery presented in the Fig.~\ref{fig:modelscompared}. The standard deviation shows that the relative width of the recovery rate distribution does not change during the recovery process.}
\label{fig:relative-strains-and-flucs}
\end{figure}

In Figure \ref{fig:relative-strains-and-flucs} the mean and the
standard deviation of spatial deformation rates are presented.
The local deformation rate is computed according to the Eq.
(\ref{eq:locstrain}). The standard deviation $\Delta \dot{\epsilon} (t)$ is defined by
 \begin{equation}
 \label{eq:local_sr_fluct}
 \Delta \dot{\epsilon} (t) =
 \sqrt{\frac{1}{N}\sum_{i,j}
 (\dot{\epsilon}^{i,j} (t) - \langle\dot{\epsilon}^{i,j}(t)\rangle)^2 }.
 \end{equation}.

The local deformation rate is evaluated at 100x73 grid points.
The decay of the averaged local deformation rate is equal to the recovery
measured from the global displacement on the image area.

The deviation from a power-law behavior at large times is probably
attributed to the
inability to measure very small strain rates and a subsequent
biasing of the average over experiments.
However, a possible physical explanation for the deviation cannot be ruled out. 
The generic features of the figure support strongly the above-mentioned
observation, that a power-law relaxation fits the data best and extends
over three decades: this then in turn should imply that the viscoelastic models 
discussed above are not valid.

\begin{figure}[th]
\includegraphics[width=.7\textwidth]{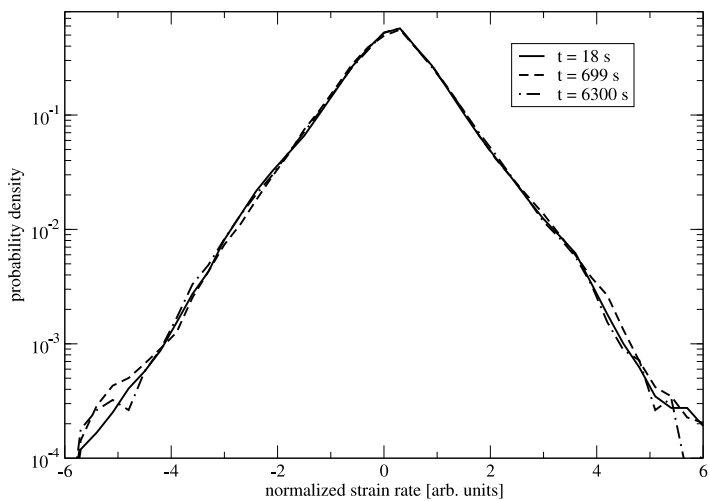}
 \caption{Normalized recovery rate distributions averaged over 32 experiments. Comparison of three different recovery times and distribution is normalized according to Eq.
(\ref{eq:normalization}). We see that the average recovery rate and its variations follow the same scaling
in time. }
\label{fig:recovery-experimens-all}
\end{figure}

\begin{figure}[th]
\includegraphics[width=.7\textwidth]{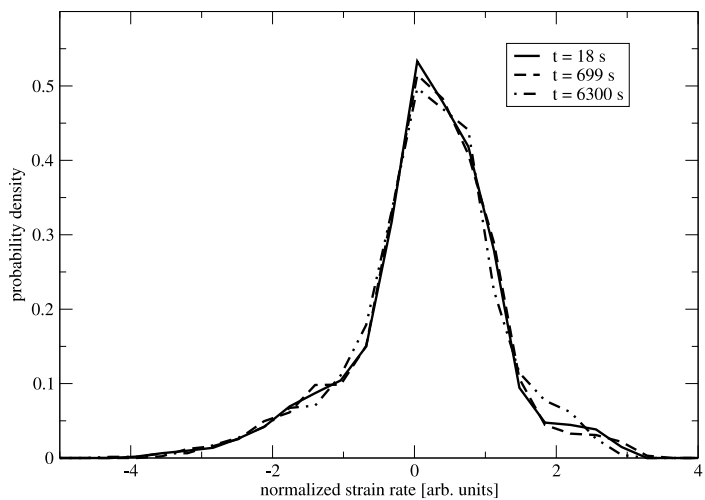}
 \caption{Normalized recovery rate distributions from a single experiment. The distribution is normalized according to Eq. (\ref{eq:normalization}). We note that the normalized form of the recovery rate distribution is unchanged during the experiment. }
\label{fig:recovery-distr-exp11}
\end{figure}

In Figure \ref{fig:recovery-experimens-all} we present
normalized distributions of local deformation rates. The deformation rate
from each experiment is normalized according to
\begin{equation}
\dot{\epsilon}' = \frac{\dot{\epsilon}-m}{s},
\label{eq:normalization}
\end{equation}

\noindent where $m$ is the average deformation rate and $s$ is the
standard deviation of the spatial deformations in the experiment.
Clearly, both the average recovery rate and its variations follow
the same scaling in time.  In Figure \ref{fig:recovery-distr-exp11}
we present as an example normalized recovery rate distributions in
one experiment. The form of the recovery rate distribution remains
unchanged during the recovery of the sample after a normalization with
$s$ as in Eq. (\ref{eq:normalization}). Note that contrary to the
previous Figure \ref{fig:recovery-experimens-all}, the data is not
presented in a semi-log scale.

For the creep recovery, the standard deviation $\Delta \dot{\epsilon}$ of the
spatial strain rates and the spatial average $\langle\dot{\epsilon}\rangle$ obey the
same scaling in time $\Delta \dot{\epsilon} \sim\langle\dot{\epsilon}\rangle\sim t^{-1.1}$.
In another study by us, on spatial and global strain rates in creep,
the fluctuations was shown to exhibit a relative
increase during Andrade's and logarithmic creep \cite{puolikas}. The
Andrade's law  was observed for the primary creep as
 $\langle\dot{\epsilon}\rangle\sim t^{-0.7}$, but the spatial fluctuations
decreased during the primary creep according to $\Delta \dot{\epsilon}
\sim t^{-0.55}$. An analogous
behavior of the ratio of creep rate fluctuations
to the average creep rate was found for logarithmic creep, 
which is observed at lower stresses.
The recovery process does not behave similarly, as the ratio of
fluctuations to the mean recovery rate is constant.
Thus, despite the fact that we observe a power law recovery the
results suggest strongly that recovery is not "inverse Andrade
creep" nor inverse logarithmic creep at the microscopic level. A direct consquence
is that the applications of superposition ideas, familiar in the theory of models of
viscoelastic response (see e.g. \cite{polymer2}) should not work as the role 
of fluctuations is different in creep and relaxation.

\subsection{Statistical properties}

\begin{figure}
\centering
\includegraphics[width=.7\textwidth]{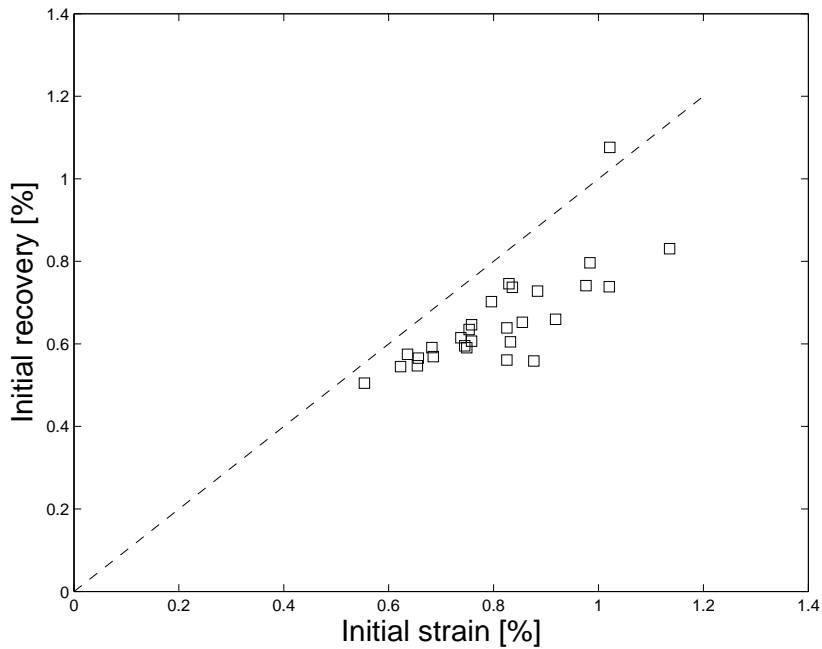}
\caption{Initial recovery as a function of initial strain. Dashed
line represents behavior of an ideally elastic material. }
\label{fig:initialstrainvsinitialrecovery}
\end{figure}

In this section we show briefly the sample to sample variation
of the recovery process.
Partially elastic behavior of paper can be observed from Figure~\ref{fig:initialstrainvsinitialrecovery} where the initial recovery
is shown to correlate with the initial strain. Here we neglect the dynamics
of the creep process during the first three seconds during the initial strain and initial recovery.
The dashed line in Figure \ref{fig:initialstrainvsinitialrecovery} represents ideal
elasticity which means that the initial recovery is equal to the initial
strain. We see that the correlation between the initial recovery and the initial strain is rather linear after the initial strain 
is over 0.6\%. By estimating where the measured data approximated by a straight line intersects 
the line of linear elasticity, one concludes that the ideal elastic behavior takes place 
when the initial strain is less than 0.5 \%. One can also consider the recovered strain at
the end of the experiment as a function of the maximum strain
at the point the creep phase is finished. It shows similar
features as Figure~\ref{fig:initialstrainvsinitialrecovery}: a roughly linear behavior for total
strains over 0.6 \%, with a slope less than one.

\subsection{Fitting a landscape model}

Next we consider an energy landscape model to look for 
interpretations of the relaxation result. The main observations
to be utilized are that we know that the strain rate scales close to
$\partial_t \epsilon \sim -1/t$, and that the local fluctuations
do not seem to become in relative terms more important with time. This means
that we may try a mean-field model  as the role of fluctuations will not increase as time progresses 
in the relaxation process. Note that coarse-grained models of crystal
plasticity \cite{morettizaiser} have difficulties in reproducing
creep and relaxation phenomena, though they include explicit elastic
interactions of the local yield strains.

Let us start with the general note that the local strain can be written as follows:
$\epsilon(x,t) = \epsilon_{pl} (x) + \epsilon_{el,comp}(x) +
\epsilon_{el,free}(x,t)$. Here, we state that at the end the load
history will leave the material with a local irreversible strain (field)
$\epsilon_{pl}$. In equilibrium (if ever reachable), this results in
internal stresses that need to be balanced, and this creates
corresponding "as such" elastic strains to maintain the
compatibility of local internal stresses: $\epsilon_{el,comp}$
\cite{ank}.
In addition, we have the leftover, the free elastic strain
$\epsilon_{el,free}$ which is a function of space and time, and it
is the one that actually relaxes away and is being measured in
the recovery. Let us consider the dynamics
of the free elastic strain in the averaged sense and drop the
dependence thereof on $x$. This means that in the absence of a
theory which accounts correctly for the spatial interactions of the
stress or strain components, we resort to a mean-field description,
where such complexity from local dynamics
is summarized with an averaged "field".

Now, the driving process in the relaxation is the fact that the free
elastic strain is not distributed equally, and in the optimal way:
the elastic energy content averaged over the sample $\langle \frac{E
\epsilon_{el}^2}{2}\rangle_x$ is higher than in the equilibrium. So
what we can write now is an equation, that states that "strain
change rate = derivative of elastic energy with respect to strain
times a typical relaxation rate". This is $\partial_t \epsilon
= -E \epsilon/\tau$. This of course would indicate a simple
exponential decay. The usual trick applied in other similar systems
with relaxation phenomena (as magnetic ones and liquid crystals \cite{clarke})
is to add a
typical barrier $U(\epsilon$), which governs the relaxation at
a given time. Thus we assume that such typical barriers at $\epsilon$
dictate the rate, and that they change with time. Then the equation reads
\begin{equation}
\partial_t \epsilon = -E \epsilon/\tau
\exp{(-U(\epsilon)/kT)}
\label{relax}
\end{equation}
as it is usually written. $kT$ has the usual meaning.

\begin{figure}[th]\
\includegraphics[width=.7\textwidth]{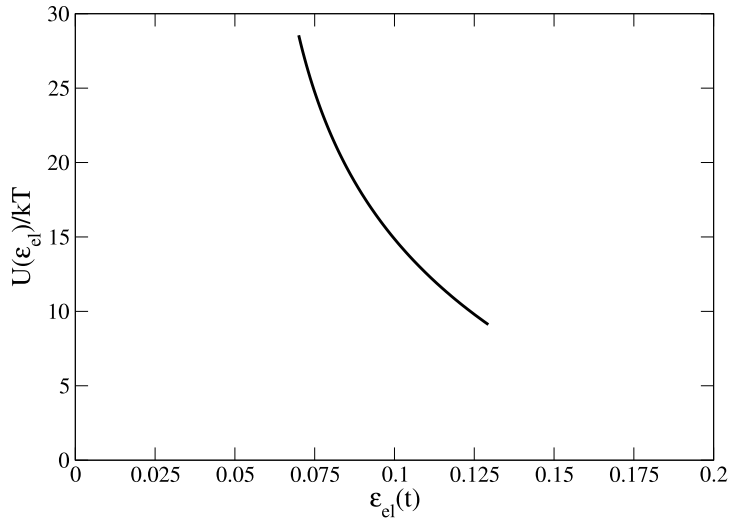}
 \caption{The result of Eq. (\ref{pote}) for the effective landscape potential $U(\epsilon)$.
 The constants have been chosen according to experiments, $\epsilon=1.3 [\%]$,
 and $A=0.58 [\%]$. }
\label{fig:pot}
\end{figure}

In our case we know that the left-hand side scales as $-1/t$.
Since $\epsilon$ is the
corresponding time derivative times $t$ plus a constant, we can just
extract $U$ (times $1/kT$) by dividing by the strain and taking a
logarithm. A constant shift factor remains ($E/\tau$) which we set to 
zero for simplicity.
We may allow for generality that $\epsilon
\sim A + Bt^{-\beta}$, whence
\begin{equation}
-U(\epsilon)/kT = \ln{\frac{\beta Bt^{-\beta-1}}{A+Bt^{-\beta}}}.
\label{pote}
\end{equation}
The important thing here is that for a reasonable $\beta$, close to
zero, the RHS goes as $\ln{(C/t)}$, where $C$ is a constant. If we
now interpret $t$ as strain solving it from the relation of time and
strain, we have in principle reconstructed the distribution of
the typical time-dependent and rate-setting barriers
in terms of strain, as given by the model fit. One can plot the
distribution and observe that such barriers get higher as the strain
relaxes, as is natural given the slowing-down of the experimentally observed relaxation, 
compared to a simple exponential decay. This is illustrated in Figure
\ref{fig:pot}, where we have utilized in the reconstruction of
$U(\epsilon)$ the parameters from the power-law fit, Eq.
(\ref{eq:pl}) from Table I. The schematic model presented here does
not take into account in any direct way the local frustration
\cite{issner} and localization of relaxation dynamics - if an
unrelaxed area is surrounded by relaxed material, this will induce
barriers to local rearrangements. The potential $U$ should however be
interpreted as a measure thereof.

The next step is to interpret the relaxation, expressed
as Eq.~(\ref{relax}) as a balance equation involving
the elastic stress $\sigma$ and internal stresses $\sigma_{int}$ is
\begin{equation}
\tau \partial_t \epsilon = -(\sigma - \sigma_{int}).
\end{equation}
The main point of the ansatz is that the relaxation stops once the internal
stresses equal the tendency to relax. Assuming that $\sigma = E \epsilon$,
one recovers the internal stresses.
In terms of the barriers, they can be written utilizing
Eq.~(\ref{relax}) again as 
\begin{equation}
\sigma_{int} = \sigma (1 - \exp(-U(\epsilon_{el})/kT)), 
\end{equation} 
which is equivalent to
\begin{equation}
	\sigma_{int} = E \epsilon (1 + \frac{\tau \partial_t \epsilon}{E \epsilon})
	\label{eq:sigmaint}
\end{equation}
for the internal stress. $\epsilon$ itself is time-dependent, and
a correction appears proportional to $\tau$ and the logarithmic
time-derivative of the strain.

\begin{figure}[th]\
\includegraphics[width=.7\textwidth]{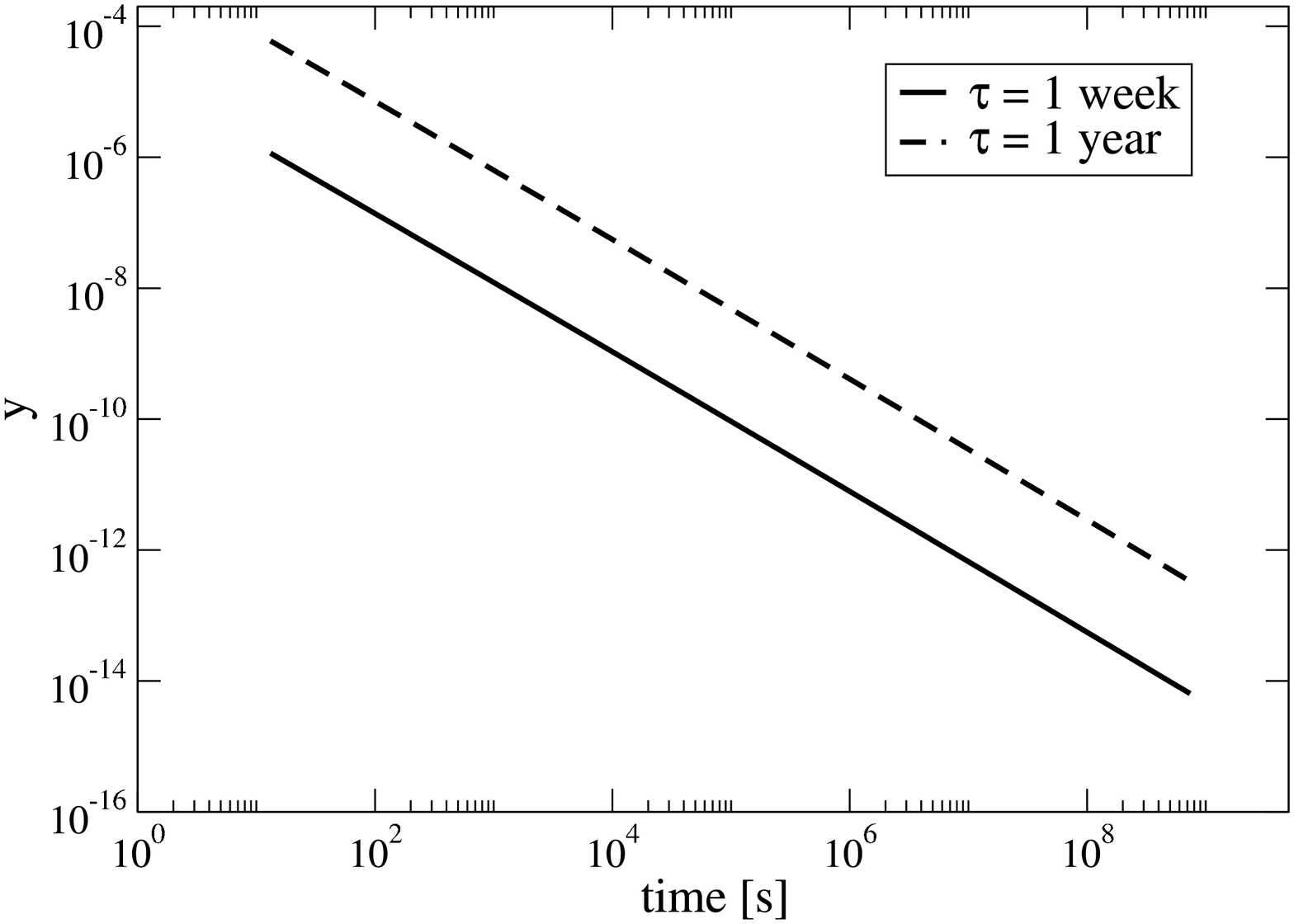}
 \caption{The expression $y = -\tau \partial_t \epsilon / E \epsilon$ is shown using 1~week and 1~year relaxation times ($\tau$). The figure depicts the effect of the barriers to the relaxation dynamics by showing the magnitude of the barrier term in equation \ref{eq:sigmaint}. The constants have been chosen according to parameters from the Table \ref{tab:fitparameters}.}
\label{fig:sigmaint}
\end{figure}

Using the power-law fit for the strain, and parameters from the 
Table \ref{tab:fitparameters}, we can estimate the contribution of the
elastic strain and the barrier term to the internal stress $\sigma_{int}$. 
One needs also to explore various guesses for $\tau$, the internal 
relaxation timescale.
To illustrate the result, we plot the correction from
the barriers to $\sigma_{int}$, or the  expression 
$y = -\tau \partial_t \epsilon / E \epsilon$ in the Figure 
\ref{fig:sigmaint}. The plot shows that if the relaxation time 
$\tau$ is of the order of week up to
years the main contribution of the relaxation of the internal 
stresses is due to the part proportional to the elastic stress. In other words, for reasonable values of $\tau$
on time-scales up from one minute, y is  always much smaller
than unity and decays in an apparent power law
fashion. This decay is in turn directly related to the
relaxation exponent $\beta$ indicated by experimental fit
to a power-law form, and in this interpretation to the
typical barrier dynamics.

The simplest implication this results in is that there are
three independent ingredients in the relaxation process: the inherent
time-scale $\tau$, the particular dynamics of internal stresses
that would be tightly coupled to the (relaxing) elastic ones,
and the slowing-down due to barriers as such.

\section{Conclusions}

We have studied, in paper creep, the recovery of the creep strain.
It turns out to be such that the phenomenon is characterized by two
features: the statistical variations in each experiment and from
sample to sample, and by the extremely slow response. The recovery
seems to follow a power-law dependence on time. The local strain rates
follow a probability distribution, which rescaled is independent of
time. This indicates the presence of collective phenomena in the
recovery dynamics.

The main technique utilized here, the DIC, is clearly an advancement
since it allows routinely to achieve good accuracy in the sample
deformation measurement, and simultaneously gives access to the
spatial strains. Thereby, the statistical behavior of local dynamics
of relaxation in a single experiment and averaged over many
experiments is available.

We have presented empirical observation on the
cross-correlations of the creep phase deformations and those of the
relaxation phase. It concerns the immediate (elastic)
relaxation and its relation to the creep strain at the beginning of
the relaxation. This has all been done in one specific case, of
material and creep test parameters, and it seems pertinent to
underline that further work would seem important here. For instance,
one should do cyclic experiments to mimic rejuvenation and aging
studies in glassy systems \cite{jpb,cloitre}.

The main results are perhaps, from the viewpoint of glassy systems,
those that regard the recovery as a function of time and
with the behavior of the distribution of local variations. 
It appears worthwhile
to study the phenomena in greater detail. The results suggest that despite
large variation in the initial creep deformation we
see that the generic power-law recovery behavior remains the same.
One feature one should pay attention to is the fact that
the recovery fluctuations scale similarly to the mean recovery
rate, in contrast to the original creep process where they
increase in relative terms \cite{puolikas}. There, it appears
that this is due to an underlying phase transition, controlled
by the applied stress.


The landscape fitting model gives way to interpret
the slow dynamics via the effective barriers that dictate 
the relaxation rate. These are not
included in viscoelastic models. Such a description makes more
clear the observation that the
relaxation is indeed of power-law form - contrary to any
viscoelastic models. The derived typical barrier increases as the
relaxation process advances, which might relate to the increasing
resistance of the surrounding, more relaxed regions to local
rearrangements. One can push the fitting to such a model further
by making an Ansatz about the role of internal stresses, which leads
to some understanding of the role of various mechanisms in the relaxation.
It would be interesting and important to develop
models that take into explicitly account the spatial part of the
dynamics, and could thus be compared with such data as the DIC
technique gives access to. One possibility might be extending
mesoscopic models of plasticity in terms of elastic
manifolds \cite{morettizaiser}.

 {\it Acknowledgements -} The authors would like to
thank for the support of the Center of Excellence -program of the
Academy of Finland, and the financial support of the European
Commissions NEST Pathfinder programme TRIGS under contract
NEST-2005-PATH-COM-043386. KCL Ltd is thanked for financial support.
We are also grateful for several constructive comments by referees..


\begin{thebibliography}{99}

\bibitem{stefano} S. Zapperi, G. Durin, \emph{The Science of Hysteresis}, vol. II, eds. G. Bertotti and I. Mayergoyz eds, Elsevier, Amsterdam, 181 (2006).

\bibitem{zaiser} M. Zaiser, Adv. Phys. {\bf 54}, 185 (2006).

\bibitem{weiss}
J. Weiss, D. Marsan, Science {\bf 299}, 89 (2003).

\bibitem{UCH-04} M. D. Uchic {et al.}, Science {\bf 305}, 986 (2004).

\bibitem{Advphys}
M.J. Alava, P.N.N.K. Nukala, S. Zapperi, Adv. Phys. {\bf 54},
347 (2006).
\bibitem{MIG-01} M.-C. Miguel { et al.}, Nature {\bf 410}, 667 (2001).

\bibitem{jpb} J.P. Bouchaud, L.F. Cugliandolo, J. Kurchan, M. M\'ezard, in: \emph{Spin-glasses and random fields}, Ed. A. P. Young, World Scientific, (1998).

\bibitem{vebook} John D. Ferry, \emph{Viscoelastic Properties of Polymers}, 3rd Ed., Wiley, (1980).

\bibitem{creep}
S. Lemerle et al., Phys. Rev. Lett. {\bf 80}, 849 (1998);
J. Koivisto, J. Rosti, M.J. Alava, Phys. Rev. Lett. {\bf 99}, 145504 (2007);
T. Nattermann, Y. Shapir, I. Vilfan, Phys. Rev. B {\bf 42}, 8577 (1990);
P. Chauve, T. Giamarchi, P. Le Doussal, Phys. Rev. {\bf B62}, 6241 (2000);
P. Le Doussal, K.J. Wiese, P. Chauve, Phys. Rev. {\bf E69}, 026112 (2004);
A. B. Kolton, A. Rosso,T.Giamarchi, Phys. Rev. Lett. {\bf 94},  047002 (2005).

\bibitem{carmen}  M. Carmen Miguel, L. Laurson, M.J. Alava,
Eur. Phys. J. {\bf B64}, 443 (2008).

\bibitem{AlavaNiskanen} M. Alava, K. Niskanen,  Rep. Prog. Phys. \textbf{69},  669 (2006).

\bibitem{santucci} S. Santucci, L. Vanel, S. Ciliberto, Phys. Rev. Lett. {\bf 93}, 095505 (2004).

\bibitem{Niskanen} K. Niskanen, \emph{Papermaking Science and Technology: Paper Physics, 2nd Ed.},
Fapet Oy, Jyv\"sskyl\"a, (2008).

\bibitem{bucky} see e.g. D.J. Hall et al., Science {\bf 320}, 507 (2008).

\bibitem{graphene} D.A. Dikin, Nature {\bf 448}, 457 (2007).

\bibitem{actine}
A.B. Bausch, K. Kroy, Nature Phys. {\bf 2}, 231 (2006);
M.L. Gardel, M.T. Valentine, J.C. Crocker, A.R. Bausch, D.A. Weitz, Phys. Rev. Lett. {\bf 91}, 158302 (2003);
R. Tharmann, M. M. A. E. Claessens, A. R. Bausch, Phys. Rev. Lett. {\bf 98},  088103 (2007).

\bibitem{polymer} R. H. Colby, L. M. Nentwich, S. R. Clingman, C. K. Ober, 
Europhys. Lett. \textbf{54}, 269 (2001)

\bibitem{polymer2} W. N. Findley, Polymer Eng. and Sci. \textbf{27}, 582 (2007).

\bibitem{duval} P. Duval, Ann. Geophys. {\bf 32}, 335 (1976).

\bibitem{Brezinski} J.P. Brezinski, Tappi Journal \textbf{39}, 116 (1956).

\bibitem{Coffin} D. W. Coffin, \emph{Advances in Paper Science and Technology}, I'Anson, S.J. (ed.),
Fundamental Research Conference, Lancashire, 651, UK, (2005).

\bibitem{hild} F.~Hild, S.~Roux, Strain \textbf{42}, 69 (2006).

\bibitem{bausch}
H.A. Bruck, Exptl. Mech. \textbf{29}, 261 (1989).

\bibitem{Lif} J.O. Lif, S. \"Ostlund, C. Fellers,  Mech. Time-Dep. Mater. \textbf{2},  245 (1999).

\bibitem{kybic} J.~Kybic, M.~Unser, IEEE Trans. Image Process. \textbf{12}, 1427 (2003).

\bibitem{sutton} M.A.~Sutton, J.L.~Turner, H.A.~Bruck, T.A.~Chae, Exp. Mech. \textbf{31}, 168 (1991).

\bibitem{kybic2}
J.~Kybic, P.~Thevenaz, A.~Nirkko, M.~Unser,
IEEE Trans. Med. Imag. {\bf 19}, 80 (2000).

\bibitem{kybicthesis} J.~Kybic, \emph{Elastic Image Registration using Parametric Deformation Models}, PhD Thesis, Ecole Polytech. Fed. Lausanne, (2001).

\bibitem{ketoja} J.A. Ketoja, J. Asikainen, S.Lehti, A.Tanaka, 
\emph{Creep of Wet Paper}, 61st Appita Ann. Conf. and Exh., Gold Coast, Australia 6-9 May, (2007).

\bibitem{puolikas} J.~Rosti et al., submitted (2009).

\bibitem{Fancey} K.S. Fancey, J. Polymer Eng. \textbf{21}, 489 (2001).

\bibitem{Fancey2} K.S. Fancey,  J. Mater. Sci. \textbf{18}, 4827 (2005).

\bibitem{Schapery2} R.A. Schapery,  Polym. Eng. Sci. \textbf{9}, 295
(1969).

\bibitem{ank}
M.J. Alava, M.E.J. Karttunen, K.J. Niskanen, Europhys. Lett. {\bf 32}, 143 (1995).

\bibitem{clarke} S.M. Clarke, E.M. Terentjev, Faraday Disc. {\bf 112}, 324 (1999).

\bibitem{issner} B.A. Isner, D.J. Lacks, Phys. Rev. Lett. {\bf 96}, 025506 (2006).

\bibitem{cloitre} M. Cloitre, R. Borrega, L. Leibler, Phys. Rev. Lett. {\bf 85}, 4819 (2000).

\bibitem{morettizaiser} 
M.~Zaiser, P.~Moretti, J. Stat. Mech. \textbf{ P08004} (2005).

\end{thebibliography}
\end{document}